\documentclass[twocolumn,twoside]{svmultivs_gm} 
\usepackage{graphicx}
\usepackage{multirow}
\usepackage{bm}
\usepackage{amssymb}

\newcommand*\aap{A\&A}

\newcommand*\aj{AJ}

\title*{Evaluate the ICRF3 axes stability via extragalactic source position time series}
\titlerunning{Evaluate ICRF3 axes stability}
\author{Niu Liu$^{1,2}$, S\'ebastien~Lambert$^3$, Felicitas Arias$^3$, Jia-Cheng Liu$^1$, Zi Zhu$^1$}
\authorrunning{Liu et al.} 
\authoremails{niu.liu@nju.edu.cn,sebastien.lambert@obspm.fr,felicitas.arias@obspm.fr,jcliu@nju.edu.cn,zhuzi@nju.edu.cn}
\institute{1. School of Astronomy and Space Science, 
Key Laboratory of Modern Astronomy and Astrophysics (Ministry of Education), Nanjing University, Nanjing 210023, P. R. China \\ 
2. School of Earth Sciences and Engineering, Nanjing University, Nanjing 210023, P. R. China \\ 
3. SYRTE, Observatoire de Paris, Université PSL, CNRS, Sorbonne Université, LNE, Paris, France}
\ContactAuthorName{Niu Liu}
\ContactAuthorTelephone{+86-025-83590036}
\ContactAuthorEmail{niu.liu@nju.edu.cn}
\NumberofInstitutions{2}
\InstitutionPostAddress{1}{N.A}
\InstitutionCountry{1}{N.A}
\InstitutionWebPage{1}{N.A}
\InstitutionPostAddress{2}{N.A}
\InstitutionCountry{2}{N.A}
\InstitutionWebPage{2}{N.A}
\begin{document}  
\maketitle       
\abstract{We present an updated study on assessing the axes stability of the third
    generation of the International Celestial Reference Frame (ICRF3)
    in terms of linear drift and scatter based on the extragalactic 
    source position time series from analyses of archival very long baseline interferometry observations. 
    Our results show that the axes of the ICRF3 are stable at a level of 10 to 20
    microseconds of arc, and it does not degrade after the adoption of the ICRF3 
    when observations from new networks are included.
    We also show that the commonly used method of deriving the position time series (four-step solution) is robust.}
%
%
\section{Introduction}
    The apparent positions of extragalactic sources are known to vary with time 
    due to their intrinsic evolution.
    This kind of astrometric instability will cause an orientation variation
    to the axes of the celestial reference frame that are defined by positions
    of extragalactic sources, a phenomenon known as the celestial frame
    instability \cite{2008A&A...481..535L}.
    Recent studies based on the extragalactic source position time series 
    measured by very long baseline interferometry (VLBI) suggest that sources
    that were considered stable are likely to become unstable as long as 
    they are observed for a longer time span \cite{2018A&A...618A..80G}.
    As a result, it is necessary to regularly monitor the astrometric behavior 
    of extragalactic sources and the axes stability of the VLBI 
    celestial frame. 

    We updated our previous work \cite{2022A&A...659A..75L} on the evaluation of the ICRF3 axes stability by extending the VLBI observations to 2022 and also by examining the robustness of the method for deriving the position time series.
    All the data and the codes (in the \textsc{PYTHON} Jupyter notebook) to reproduce the results of this work can be accessed publicly online\footnote{\url{https://git.nju.edu.cn/neo/icrf3-axis-stability-2022}}.
\section{Data}
    We used observations in 7146 VLBI regular sessions made between November 1979 and December 2021 that are publicly available at the data center\footnote{\url{ftp://ivsopar.obspm.fr/vlbi/ivsdata/vgosdb}} of the International VLBI Service for Geodesy and Astrometry \cite{2017JGeod..91..711N}.
    These data were processed with the Calc/Solve software \cite{1986AJ.....92.1020M} in the global solution mode.
    
    To produce the position time series for each sources, the observed sources were divided into $N$ subsets in a manner so that each subset contains same amount of the ICRF3 defining sources and non-defining sources and has an as identical and uniform as possible sky distribution.
    Then, $N$ separate VLBI global solutions were carried out, which treated positions of sources in one respective subset as the sessionwise parameters and thus provided the position time series for these sources.
    These $N$ solutions together provided the position time series for the full source sample.
    We may call this method of producing the source position time series as ``$N$-step'' method.
    Different values of $N$ were used by various authors; for example, $N=4$ in \cite{2013A&A...553A.122L} and $N=10$ in \cite{2018A&A...618A..80G}.
    In this work, we set $N$ to 4, 8, 12, 16, and 20 in order to test the robustness of the ``$N$-steps'' method in deriving the position time series and also estimate the influence of the different choices of $N$ on the evaluation of the ICRF3 axes stability. 
    
    Finally, we obtained position time series for 6032 extragalactic sources, including all 303 ICRF3 defining sources, which were considered in this work.
    The median number of observed sessions and median of mean observing epoch for the ICRF3 defining sources was 157 and 2013.5, respectively.
    The observation time span ranged from 4.13~yr to 42.06~yr.

\section{Analysis}
    We adopted the same methods as used in \cite{2022A&A...659A..75L} to evaluate the stability of the ICRF3 axes.
    On the one hand, we estimated the spin of ICRF3, that is, the change rate of the orientation of the ICRF3 axes, through the apparent proper motion field of extragalactic sources, the latter being derived from the position time series.
    The value of spin multiplied by the length of the observation period offered an estimate of the stability of the ICRF3 axes.
    On the other hand, we constructed yearly representations of ICRF3 by annually averaging the source positions within a one-year window and studied the temporal variation in the axes orientation of these yearly celestial frames, which assessed the stability of the ICRF3 axes from another aspect.
    
\subsection{Spin of ICRF3}

\begin{figure*}[htb!]
\includegraphics[width=.5\textwidth]{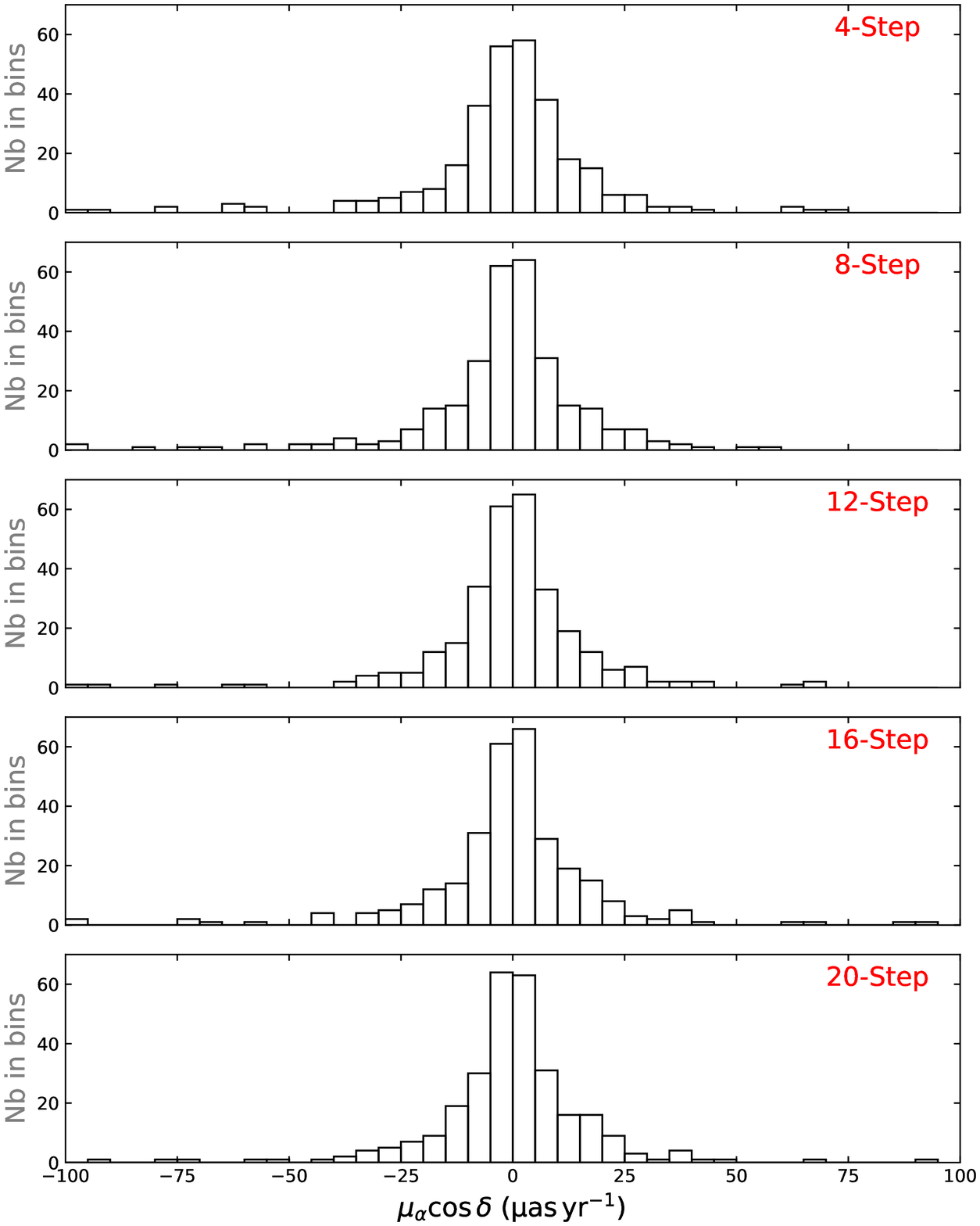}
\includegraphics[width=.5\textwidth]{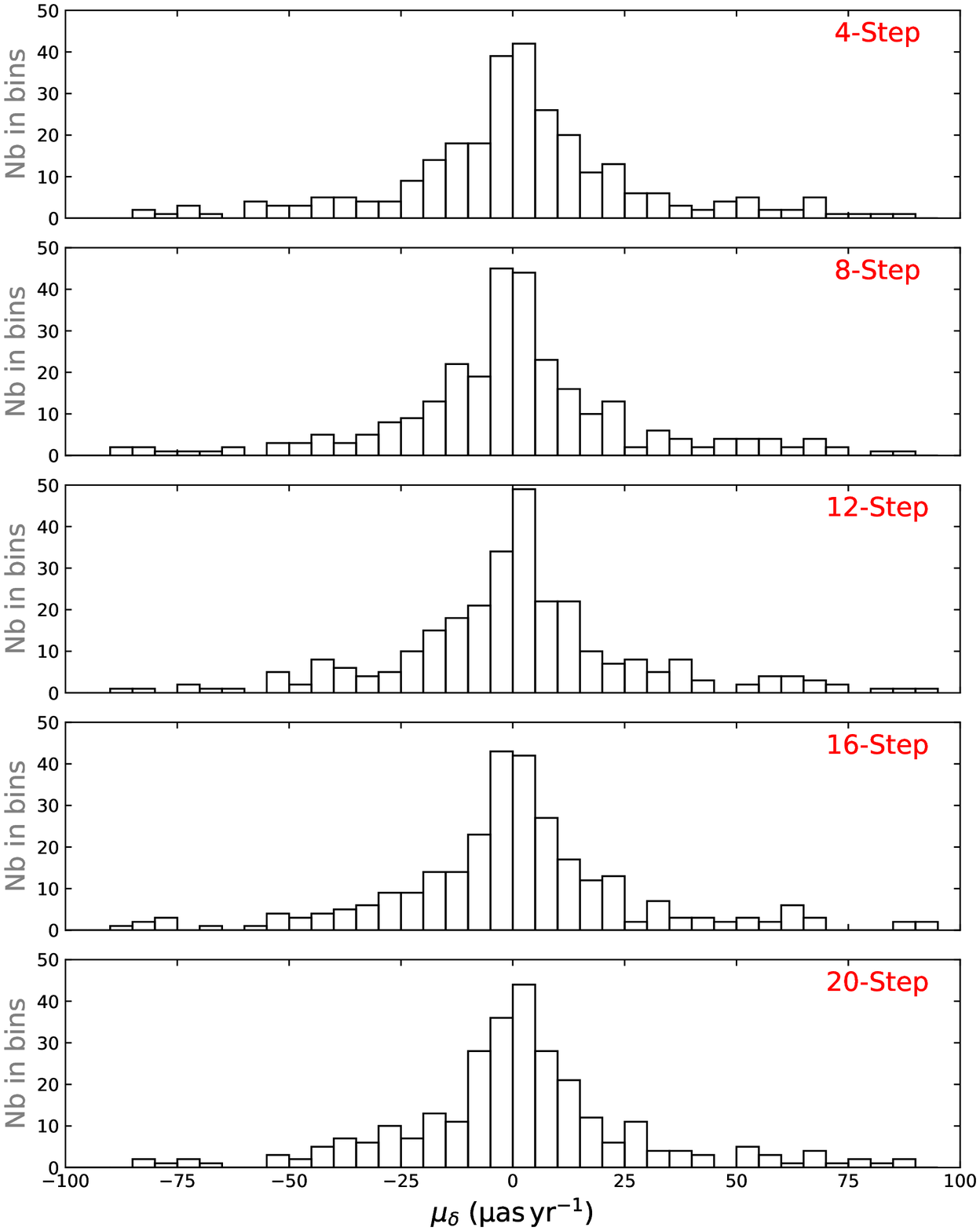}
\caption{Distributions of apparent proper motions in right ascension (\textit{left}) and declination (\textit{right}) for the ICRF3 defining sources based on position time series from different solutions.}
\label{fig:apm-dist}
\end{figure*}

\begin{figure}[htb!]
\includegraphics[width=.5\textwidth]{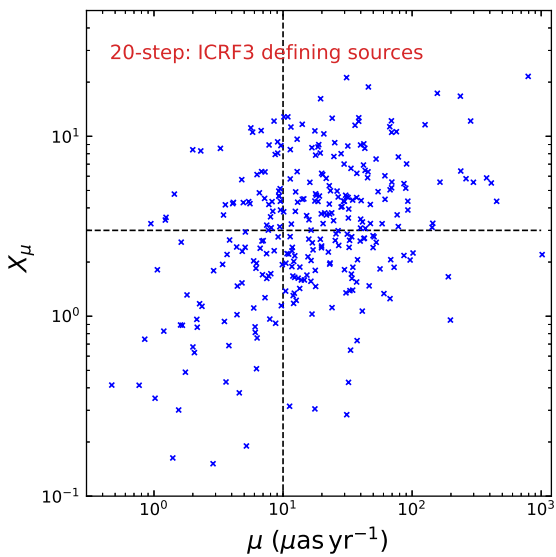}
\caption{Distribution of total apparent proper motion against their significance from the 20-step solution.}
\label{fig:apm-sig-dist}
\end{figure}
\begin{figure}[htb!]
\includegraphics[width=.5\textwidth]{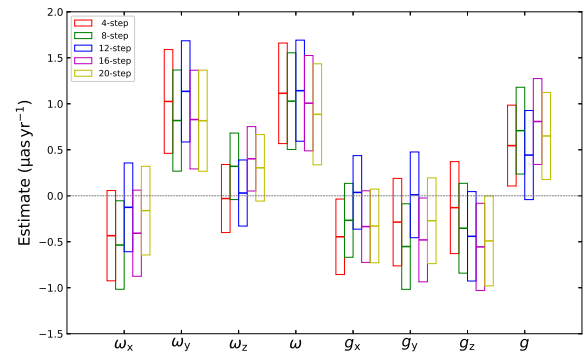}
\caption{Estimate of the spin and glide parameters.}
\label{fig:vsh01}
\end{figure}
\begin{figure*}[htb!]
\includegraphics[width=.5\textwidth]{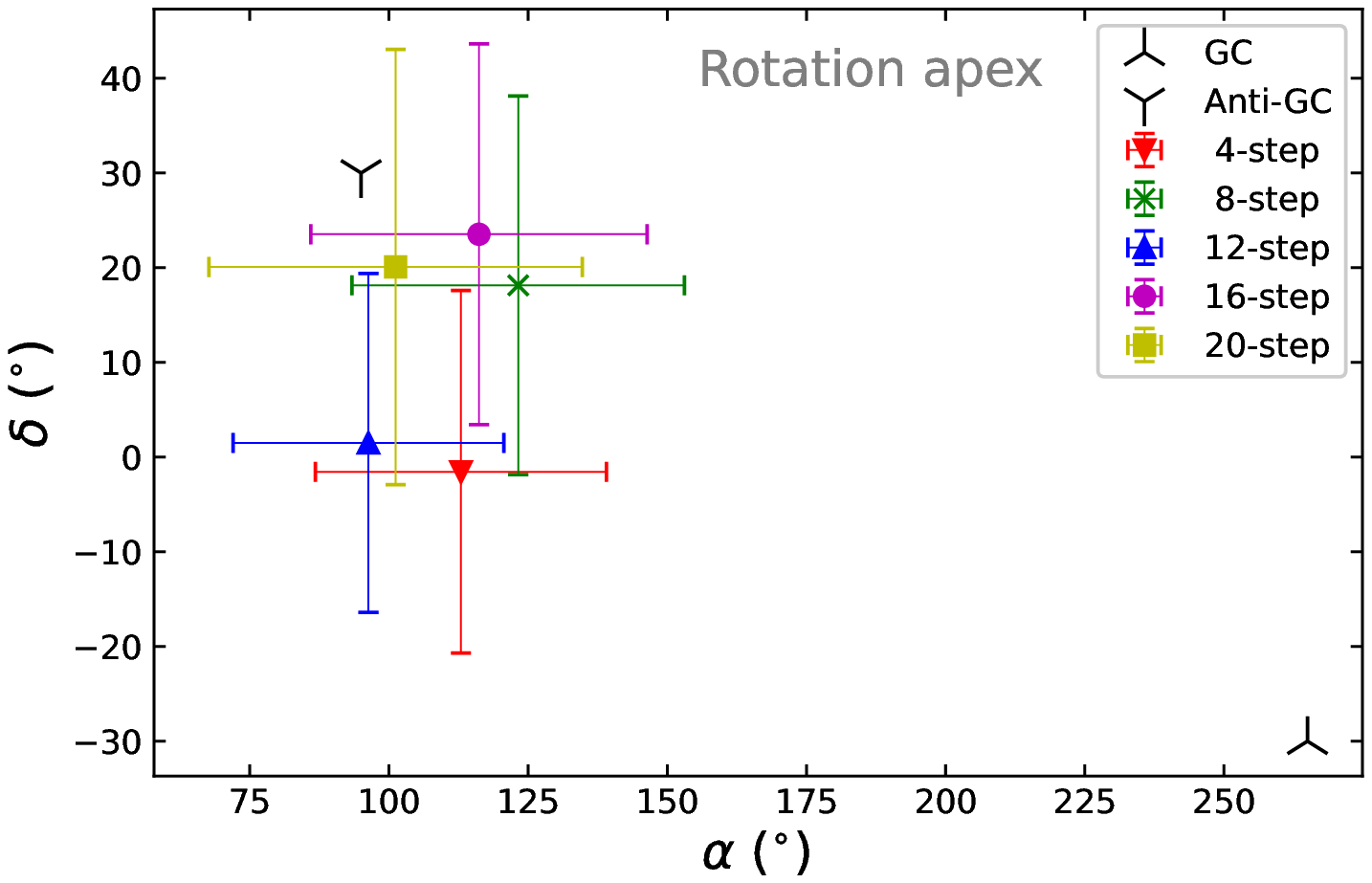}
\includegraphics[width=.5\textwidth]{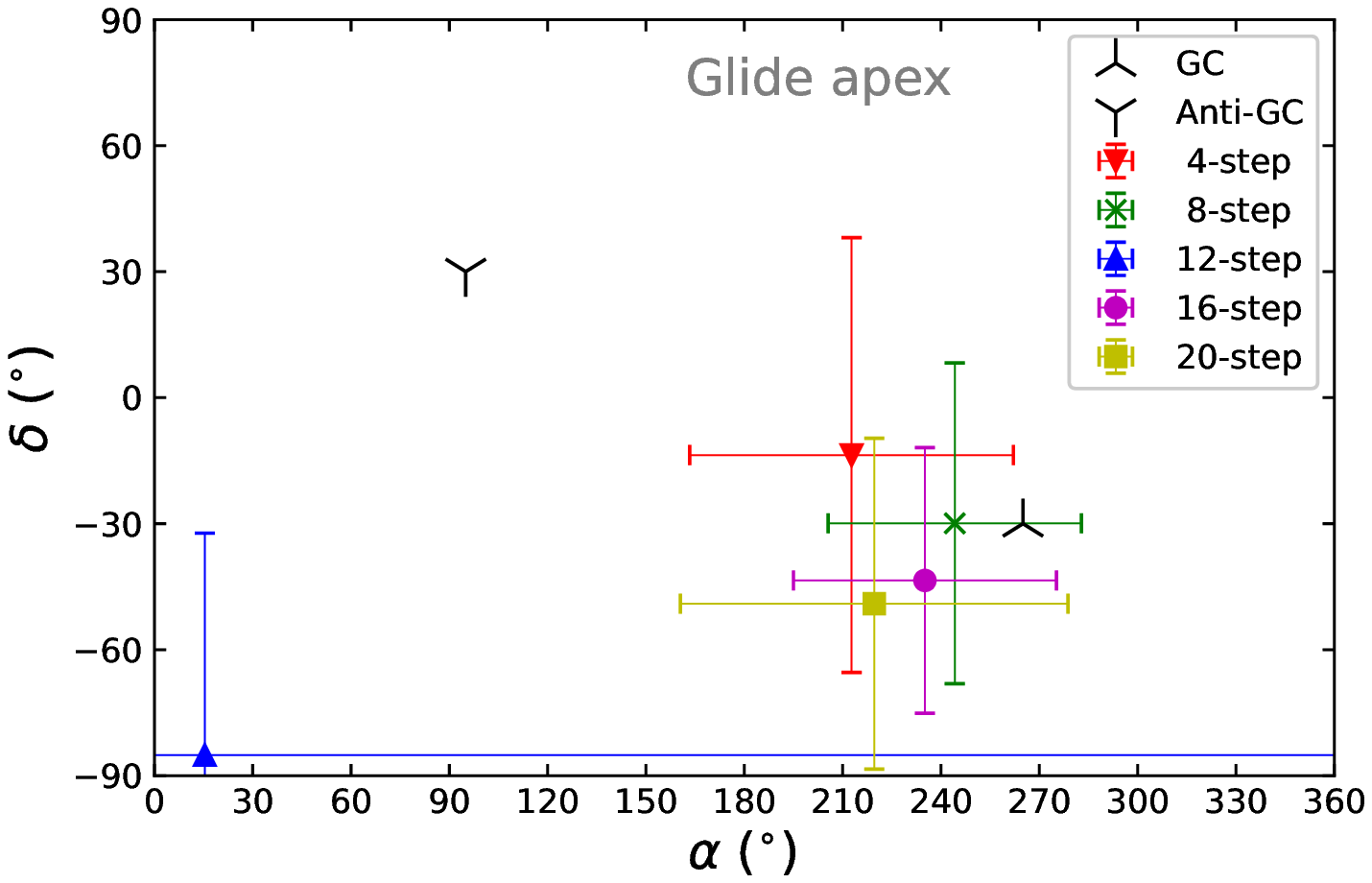}
\caption{Location of the apexes of the spin (\textit{left}) and glide (\textit{right}) vectors.}
\label{fig:rot-glide-apex}
\end{figure*}

    The apparent proper motion was estimated by the least-squares fitting to a linear motion model as
\begin{equation}
    \left[ \begin{array}{c}
         \alpha(t)   \\
         \delta(t) 
    \end{array}\right] = 
    \left[ \begin{array}{c}
         \mu_\alpha   \\
         \mu_\delta 
    \end{array}\right] (t-t_0) +
    \left[ \begin{array}{c}
         \alpha(t_0)   \\
         \delta(t_0) 
    \end{array}\right],
\end{equation}
    where $t_0$ was set to be the mean observing epoch.
    Data points whose distances to the mean position being three times greater than the corresponding uncertainties were removed from the position time series before fitting.
    The inverse of the full covariance matrix of each data point position time series was used as the weight.
    Figure~\ref{fig:apm-dist} presents the distributions of the apparent proper motion for the ICRF3 defining sources based on different position time series solutions, which noticeable differences can be found in the declination component.
    Nearly half of the sources show an apparent proper motion greater than $\mathrm{30\,\mu as\,yr^{-1}}$ in either right ascension or declination.
    To examine the statistical significance of these apparent proper motions, we plotted the distribution of the total apparent proper motion $\mu$ and its significance $X_{\mu}$ from the 20-step solution as an example in Fig.~\ref{fig:apm-sig-dist},
    where limits of $\mu\,\le\,\mathrm{10\,\mu as\,yr^{-1}}$ and $X_{\mu}\,\le\,3$ filter out approximately 20\% of sources.
    The source locating in the upper right corner with $\mu\,\simeq\,\mathrm{0.8\,mas\,yr^{-1}}$ and $X_{\mu}\,\simeq\,21.5$ is 2220$-$351, which may need to be studied further.
    
    We modeled the apparent proper motion filed by the first degree of the vector spherical harmonics (VSH) \cite{2012A&A...547A..59M} as 
\begin{equation} \label{eq:vsh01}
    \begin{array}{rl}
        \mu_{\alpha}\cos\delta  = &-\omega_{\rm x}\cos\alpha\sin\delta  - \omega_{\rm y}\sin\alpha\sin\delta + \omega_{\rm z}\cos\delta \\
                    &-g_{\rm x}\sin\alpha + g_{\rm y}\cos\alpha, \\
        \mu_{\delta}    = &+\omega_{\rm x}\sin\alpha - \omega_{\rm y}\cos\alpha \\
                    &-g_{\rm x}\cos\alpha\sin\delta  - g_{\rm y}\sin\alpha\sin\delta + g_{\rm z}\cos\delta, \\
    \end{array}
\end{equation}
    where $\bm{\omega}=(\omega_{\rm x}, \omega_{\rm y}, \omega_{\rm z})^{\rm T}$ stands for the spin of the celestial frame and $\bm{g}=(g_{\rm x}, g_{\rm y}, g_{\rm z})^{\rm T}$ represents the dipolar pattern due to, for example, the Galactic aberration effect \cite{2020AcASn..61...10L}.
    These parameters were estimated via a least-square fitting to the apparent proper motions of all the ICRF3 defining sources weighted by the inverse of their covariance matrix, whose results are shown in Fig.~\ref{fig:vsh01}.
    Different position time series gave consistent results, that is, only the spin around the Y-axis deviated from zero by approximately $\mathrm{0.4\,\mu as\,yr^{-1}}$ to $\mathrm{0.5\,\mu as\,yr^{-1}}$ considering the associated formal uncertainties.
    We also determined the spin parameters from a bootstrap resampling analysis based on 1000 randomly picked samples to obtain a realistic estimate of the parameter uncertainties, which were found to be approximately $\mathrm{0.7\,\mu as\,yr^{-1}}$.
    Figure~\ref{fig:rot-glide-apex} displays the apex locations of the spin and glide vectors, which are close to the directions of the anti-Galactic center and Galactic center, respectively.
    The possible explanation is that there is some residual Galactic aberration effect in the ICRF3, which requires further investigation.
    
    Considering the observation period of approximately 42\,yr and the nonzero deviation of $\mathrm{0.4\,\mu as\,yr^{-1}}$--$\mathrm{0.5\,\mu as\,yr^{-1}}$ for the spin of ICRF3, the accumulated deformation in the direction of the ICRF3 axes is on the order of $\mathrm{10\,\mu as}$--$\mathrm{20\,\mu as}$.

\subsection{Variation in the ICRF3 axes orientations}

\begin{figure*}[htb!]
\includegraphics[width=.5\textwidth]{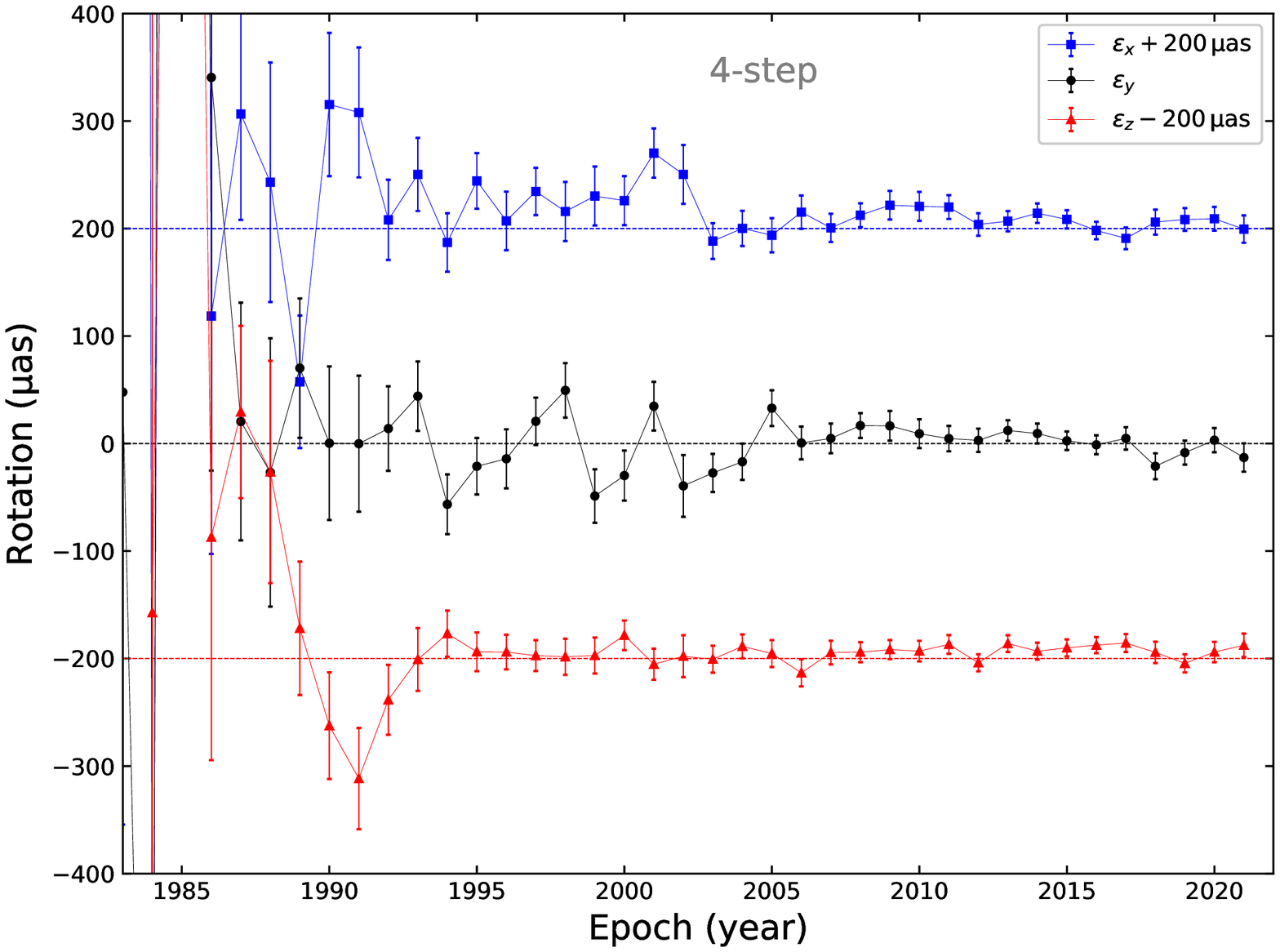}
\includegraphics[width=.5\textwidth]{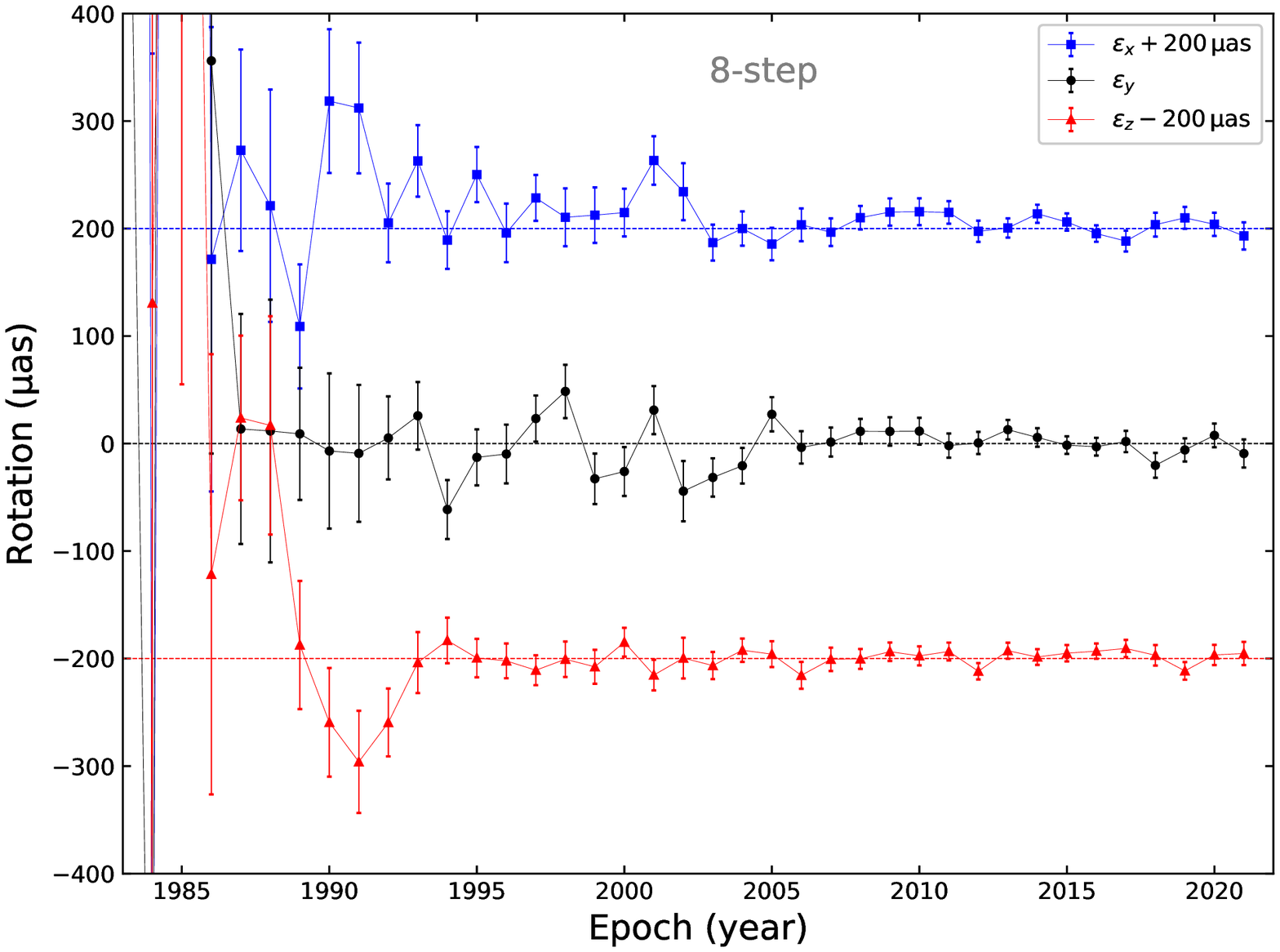}
\includegraphics[width=.5\textwidth]{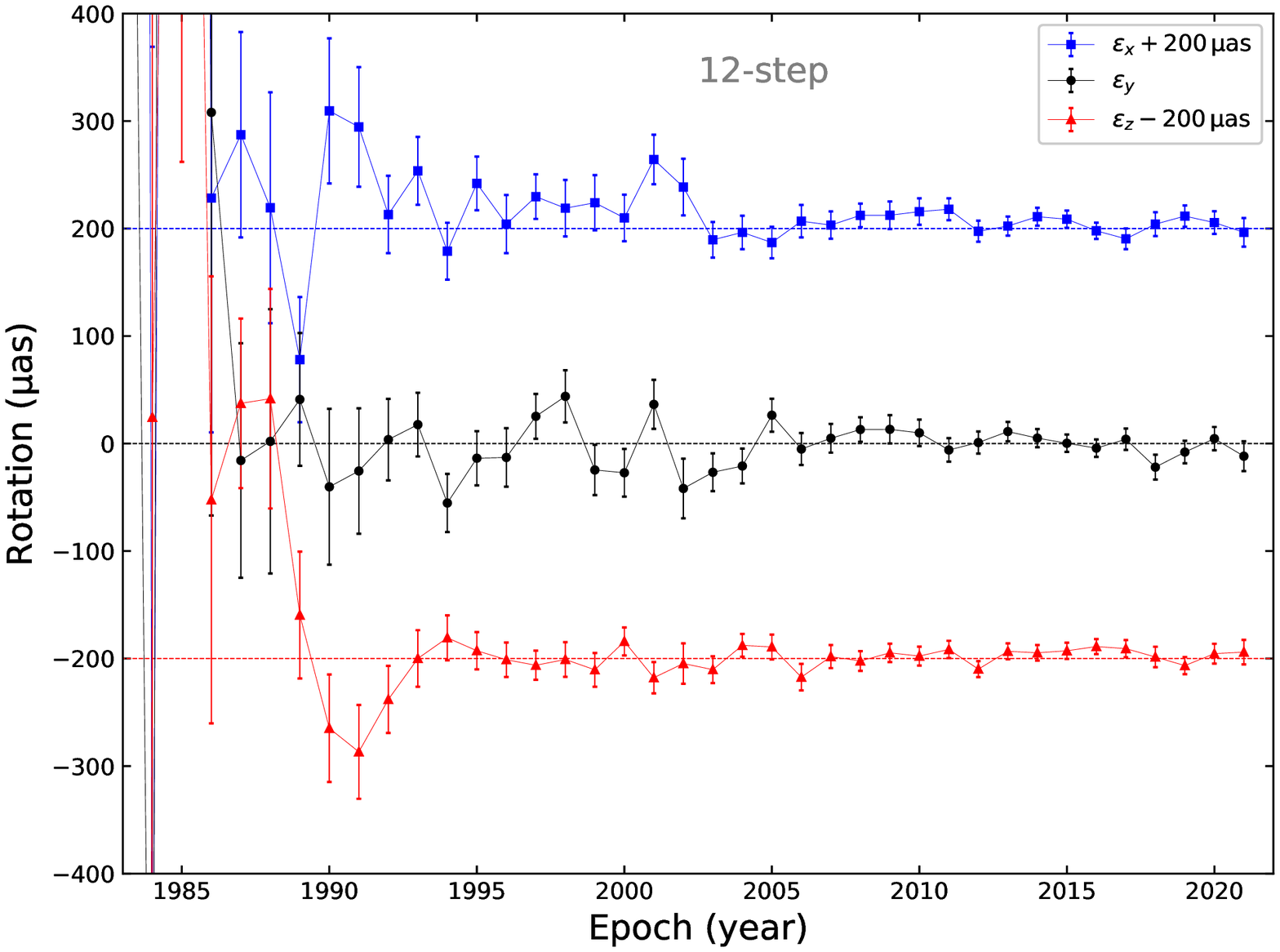}
\includegraphics[width=.5\textwidth]{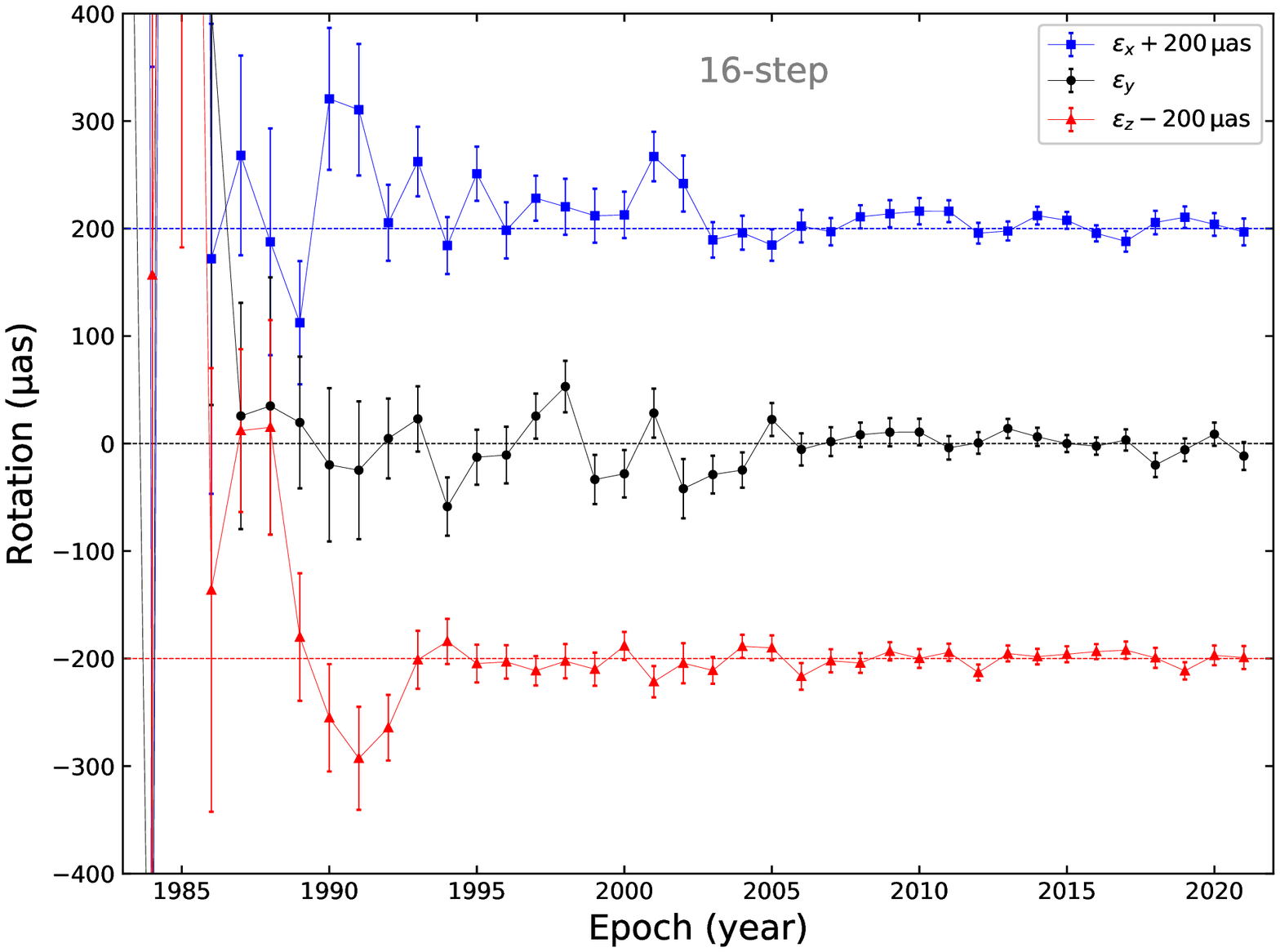}
\includegraphics[width=.5\textwidth]{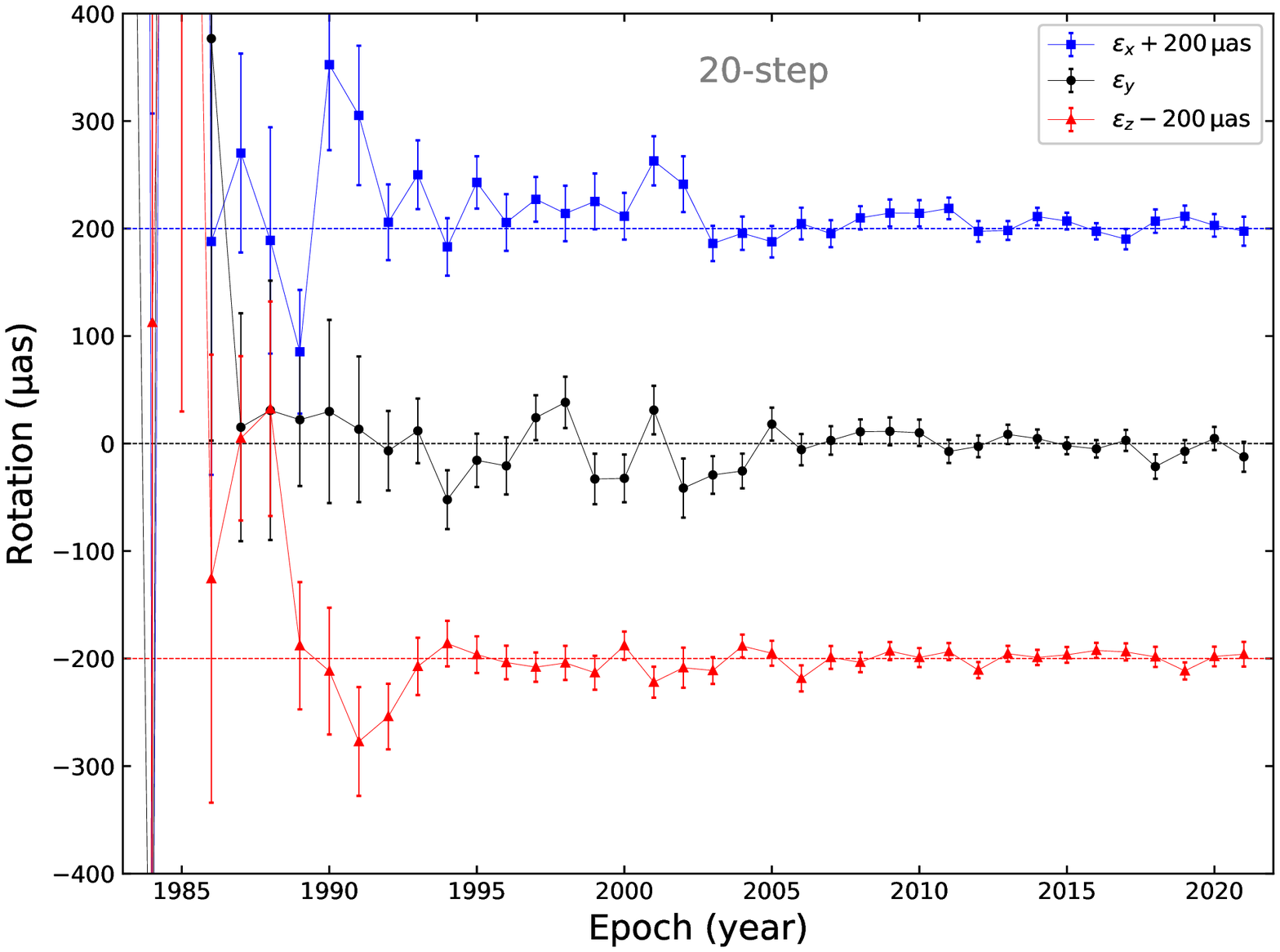}
\caption{Orientation offsets of yearly celestial frames with respect to the ICRF3.}
\label{fig:orient-var}
\end{figure*}

    The annual representations of ICRF3 were constructed by the mean positions of the ICRF3 defining sources within each one-year observing window.
    The orientation offsets of these yearly celestial reference frames with respect to ICRF3 were estimated by a weighted least-squares fitting of the first degree of the VSH that is similar to Eq.~(\ref{eq:vsh01}).
    Figure~\ref{fig:orient-var} displays the temporal variations of the orientation of the yearly celestial frame axes.
    Different position time series solutions gave consistent results.
    We computed the weighted root-mean-squares (WRMS) of the orientation offsets and tabulated them in Table~\ref{tab:orient-var}.
    The WRMS is approximately $\mathrm{10\,\mu as}$--$\mathrm{20\,\mu as}$ over the whole observation span, and it is reduced to no greater than $\mathrm{10\,\mu as}$ if only considering post-2019 data, that is, the adoption of the ICRF3 as the fundamental celestial reference frame.

\begin{table}[htb!]
\caption{Weighted root-mean-squares of orientation offsets of the yearly celestial reference frames with respect to ICRF3 in the unit of microsecond of arc.}
\begin{tabular}{| r | c | c | c | c | c | c | c | c | c |} 
\hline
\multirow{2}*{Solution} &\multicolumn{3}{|c|}{1979 -- 2022} &\multicolumn{3}{|c|}{1995 -- 2022} &\multicolumn{3}{|c|}{2018 -- 2022} \\
\cline{2-4} \cline{5-7} \cline{8-10}
& $\epsilon_{\rm x}$   & $\epsilon_{\rm y}$ &$\epsilon_{\rm z}$  & $\epsilon_{\rm x}$   & $\epsilon_{\rm y}$ &$\epsilon_{\rm z}$  & $\epsilon_{\rm x}$   & $\epsilon_{\rm y}$ &$\epsilon_{\rm z}$  \\
\hline
 4-step  &18    &19    &12    &13    &15    & 7    & 4    & 7    & 7 \\
 8-step  &16    &17    &12    &12    &13    & 7    & 7    & 7    & 8 \\
12-step  &16    &17    &13    &11    &13    & 8    & 6    & 7    & 6 \\
16-step  &17    &17    &12    &12    &14    & 8    & 5    & 8    & 7 \\
20-step  &16    &16    &12    &12    &13    & 8    & 6    & 7    & 7 \\
\hline
\end{tabular}
\label{tab:orient-var}
\end{table}

\section{Conclusions}
    We evaluate the stability of the ICRF3 axes based on the VLBI observations between November 1979 and December 2021, which is found to be at $\mathrm{10\,\mu as}$--$\mathrm{20\,\mu as}$.
    We also find that the commonly used four-step method of deriving the position time series is sufficiently robust.

\section*{Acknowledgements}

This work is supported by the National Natural Science Foundation of China (NSFC) under grant Nos. 11833004 and 12103026 and by the China Postdoctoral Science Foundation (grant No. 2021M691530).

\end{document}